\documentclass[twocolumn,showpacs,preprintnumbers,amsmath,amssymb,superscriptaddress]{revtex4}
\usepackage{amsmath}
\usepackage{amssymb}

\input epsf
\input rotate

\def\infintd3r{ \int_{-\infty}^\infty d^3r\,}
\def\intd3r{ \int d^3r\,}

\def\laplace1d{\frac{d^2}{dx^2}}
\def\plaplace1d{\frac{d^2}{d{x'}^2}}

\def\padr2{\frac{\partial^2}{\partial r^2}}

\def\bea{\begin{eqnarray}}
\def\eea{\end{eqnarray}}
\def\ben{\begin{equation}}
\def\een{\end{equation}}
\def\benu{\begin{enumerate}}
\def\enu{\end{enumerate}}

\def\sss{\scriptscriptstyle\rm}

\def\1var{(\bx_1...\bx\N)}

\def\bx{{x}}

\def\N{_{\sss N}}

\def\sph_int{ {\int d^3 r}}

\begin{document}

%\preprint{RUTGERS DFT GROUP: preprint WMB02}

\title{{Rydberg transition frequencies from the Local Density Approximation}}
%\title{{A density functional approach to low-energy elastic
%electron-ion scattering}}
\author{Adam Wasserman}
\affiliation{Department of Chemistry and Chemical Biology, Rutgers University, 610 Taylor Rd., Piscataway, NJ 08854-8087, USA}
\author{Kieron Burke}
\affiliation{Department of Chemistry and Chemical Biology, Rutgers University, 610 Taylor Rd., Piscataway, NJ 08854-8087, USA}

\date{\today}
\begin{abstract}
A method is given that extracts accurate Rydberg excitations from 
LDA density functional calculations, despite the short-ranged potential. For the case of He and Ne,
the asymptotic quantum defects predicted by LDA are in less than
5\% error, yielding transition frequency errors of less than $0.1$eV.
\end{abstract}

\maketitle

%\section{Introduction}

The excitation energy spectrum of atoms, molecules, clusters, and
solids can now be accurately calculated via
Time-dependent Density Functional Theory (TDDFT) \cite{RG84}. In
Casida's matrix formulation \cite{C96}, first the self-consistent
solution of the ground-state Kohn-Sham (KS) equations is found
\cite{KS65}, and the differences between occupied and unoccupied KS
orbital energies may then be regarded as a first approximation to the
true excitations of the system. In a second step, these KS frequencies
are corrected to become the true transitions of the many-body
system. The quality of the results depends crucially on the functional
employed for the solution of the ground-state problem.

The Local Density Approximation (LDA) is the simplest and historically
most successful approximation in DFT \cite{KS65}. Whereas new
generations of functionals have achieved better accuracy than LDA for
many properties, its ratio of reliability to simplicity has no
paragon. But a well-known shortcoming of the ground-state LDA
potential of an atom or molecule, already recognized by Tong and Sham
in the early days of the LDA \cite{TS66}, is that it decays
exponentially at large distances, rather than as $-1/r$ as the exact
KS potential does. As a consequence, the LDA potential does not
support a Rydberg series of bound states. Also, the magnitude of the
highest occupied molecular orbital (HOMO) is typically too small by
several eV in an LDA calculation, so HOMO $\to$ Rydberg transitions
appear as HOMO $\to$ continuum excitations. Because of this
limitation, several schemes have been devised to asymptotically
correct the LDA potentials. Casida and Salahub (CS) \cite{CS00} proposed a
``shift-and-splice'' approach consisting on shifting the LDA potential
downwards in the bulk regions and joining it continuously with the van
Leeuwen-Baerends potential \cite{LB94} where they cross. Tozer and
Handy (TH) proposed a very similar procedure \cite{TH98} using a
Fermi-Amaldi tail as suggested by Zhao, Morrison and Parr
\cite{ZMP94}, and a more sophisticated way to smoothen the transition
from the inner to the outer potential. Wu, Ayers, and Yang (WAY)
\cite{WAY03}, noting that neither of these potentials yield an energy
minimum, proposed a variational method for correcting the
exchange-correlation (xc) potential. Their construction imposes the
correct asymptotic behavior by using the Fermi-Amaldi form as a fixed
reference potential, and the coefficients of a linear combination of
basis functions are determined through minimization of the energy for
a given choice of energy functional. Their asymptotically corrected
LDA potentials very closely resemble the pure LDA ones shifted
downwards in energetically important regions. For any finite basis
set, however, they are no longer functional derivatives of the LDA
energy functional. Interestingly, the results obtained so far for
excitation energies with the WAY method are not as good
as those obtained with TH potentials that retain the pure
LDA form in the core regions \cite{WCY05}. All of these methods
have improved upon LDA on the prediction of Rydberg excitation frequencies,
leading to the conclusion that the correct long-range behavior of the
potentials is indispensable to describe such excitations. But we show
here that in fact {\em it is not}.

The purpose of the present work is to show how the Rydberg excitation
energies are encoded in the short-ranged LDA potentials. Imposing the
correct asymptotic behavior is one way to decode them, but not the
best one, since different tails lead to different answers (e.g. van
Leeuwen-Baerends {\em vs.} Fermi-Amaldi), obscuring the information
provided by a {\em pure} LDA calculation.

It was recently shown \cite{WMB03} that in spite of incorrectly
describing photoabsorption as if it were photoionization, the
oscillator strengths of Rydberg excitations show up in the LDA spectrum
as continuum contributions with excellent optical intensity. The
dipole matrix elements for HOMO $\to$ Rydberg transitions are accurate
in LDA because (1) the shape of the LDA HOMO is very close to that of
the exact KS HOMO, even if its energy is not, and (2) LDA continuum
orbitals at frequencies corresponding to HOMO $\to$ Rydberg
transitions, are also very close to the exact KS Rydberg orbitals in
the crucial region for optical absorption, i.e., where the HOMO has
high amplitude. The underlying physical reason for this is simple
\cite{WMB03}: the LDA xc-potential is very close to the exact
xc-potential near the nuclei, and runs almost parallel to it in the
valence regions \cite{P85}.

An obvious objection to this claim of success of the LDA is that
it cannot predict the {\em positions} of the Rydberg excitations,
even if it produces an ionization envelope that approximates well
the discrete photoabsorption spectrum. We now show that, in fact,
LDA {\em does} predict the position of high-$n$ Rydberg excitations
very accurately. We use concepts of quantum defect theory,
developed before the advent of DFT by Ham \cite{H55} and Seaton
\cite{S58}. The quantum defect $\mu_{nl}$ parametrizes the energy
$E_{nl}$ of a Rydberg state as: \ben
E_{nl}=-\frac{1}{2(n-\mu_{nl})^2}~~. \label{qd} \een For an
electron orbiting in a Coulomb field outside an ionic core, as in
a high-$n$ Rydberg state, $\mu_{nl}$ represents the effect of the
field that prevails {\em within} the core. Although the Coulomb
field outside the core is invoked for its definition in
Eq.(\ref{qd}), the actual number $\mu_{nl}$ is determined only by
the forces within \cite{F81}. It is typically a very
smooth function of $n$, and approaches rapidly the asymptotic quantum defect, $\mu_{\infty l}$, as $n\to\infty$.

We will focus on the KS asymptotic quantum defect of the
$(l=0)$-Rydberg series that converges to the first ionization
threshold of an atom (the $l=0$ subscript will be dropped from now
on). We first propose a method to extract $\mu_n$ from a given
orbital, and illustrate it with a simple example. This method was
inspired by Fano's original discussion in Ref.\cite{F81}. We then
apply it to the cases of He and Ne, where the exact KS quantum defects
are known, and show that the LDA produces $\mu_{\infty}$'s which are
in less than 5\% error.
%conclude that this new victory of
%LDA is not surprising, considering that the LDA potential is known to
%describe well the forces within the core, where $\mu$ is determined.
%
%\subsection{Quantum defect from its orbital}

{\underline{\em {Quantum defect from an orbital}}}:
Consider a long-range potential that equals $-1/r$ for $r\geq r_0$. The
solution of the radial Schr\"{o}dinger equation is well known for
$r\geq r_0$. For negative energies $E<0$, the physically
acceptable solutions are Whittaker functions (we will restrict the analysis to $s$-states):
\ben
\phi_{>r_0}(r)=AW_{1/k,1/2}(2kr) ~~,\een where $A$ is a constant
and $k=\sqrt{2|E|}$. The logarithmic derivative of $\phi_{>r_0}$
is given by: \ben \frac{d\ln
\phi_{>r_0}}{dr}=\frac{1}{n^*}-\frac{n^*}{r}-\frac{1}{r}\frac{U(-n^*;2;2r/n^*)}{U(1-n^*;2;2r/n^*)}
\label{ld}\een Here $k$ was written as $k=(n^*)^{-1}$, with $n^*=(n-\mu_n)$, where $n$
numbers the bound state, and $\mu_n$ is the quantum defect; $U$ is the
confluent hypergeometric function \cite{AS70}.

Regardless of the shape of the potential for $r<r_0$, the logarithmic
derivative of $\phi_{<r_0}$ must equal that
of $\phi_{>r_0}$ at $r_0$. Now suppose that an orbital is given to us, with the information that it is the $n=8$ state of a potential
that possesses a Coulomb tail. We can immediately obtain $\mu$ from this orbital by solving
Eq.(\ref{ld}) numerically, using $n=8$ and some large value of $r$.
If we observe that $\mu(r)$ changes as $r$ is increased, we can conclude that
Eq.(\ref{ld}) is being used in the region where $r<r_0$, and its solution cannot be interpreted as the quantum defect.

\begin{figure}
\begin{center}
\epsfxsize=80mm \epsfbox{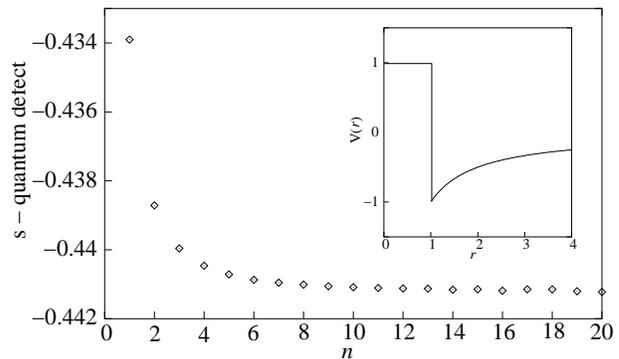}
\caption{$s$-quantum defect as a function of $n$
for the potential shown in the inset. The quantum defect is a smooth
function of $n$ and converges rapidly to its asymptotic value,
$\mu_{\infty}=-0.441$.} \label{f:qdvsn_example1}
\end{center}
\end{figure}

\begin{figure}
\begin{center}
\epsfxsize=85mm
\epsfbox{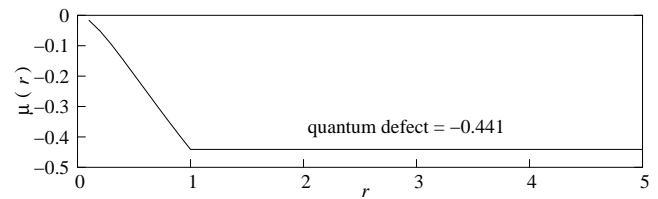}
\caption{Solution of Eq.(\ref{ld}) for $\mu$ as a function of $r$, for the potential of Fig.\ref{f:qdvsn_example1}; the logarithmic derivative of the $n=20$ orbital was found numerically as a function of $r$, and at each location $r$, it was inserted into Eq.(\ref{ld}) to get $\mu(r)$.}
\label{f:qd_example1}
\end{center}
\end{figure}

For example,
%\subsection{Two examples}
%\subsubsection*{Potential with constant core and Coulomb tail}
consider a potential which is equal to a constant $C$ for $r<r_0$
and to $-1/r$ for $r\geq r_0$. For $r_0=1$ the matching condition
is: \ben
\tilde{k}\coth\tilde{k}=\frac{1}{n^*}-n^*-\frac{U(-n^*;2;2/n^*)}{U(1-n^*;2;2/n^*)}\label{ld1}\een
where $\tilde{k}=\sqrt{2|E-C|}$. Eq.(\ref{ld1}) was solved for the
first 20 bound states. For $C=r_0=1$ the asymptotic quantum defect
is $\mu_{\infty}=-0.441$, see Fig.\ref{f:qdvsn_example1}. Figure
\ref{f:qd_example1} shows $\mu(r)$, the solution of
Eq.(\ref{ld}) as a function of $r$, for the $n=20$ orbital. Clearly, the
quantum defect for a given state can be obtained by looking at the
respective orbital anywhere in the region $r>r_0$. In particular,
it can be obtained {\em at} $r_0$. It represents the accumulation
of phase due to the non-Coulombic potential in the region of
$r<r_0$ up to $r=r_0$.

Imagine now that the potential is altered by truncating the Coulomb
tail far away, at $r_1>>r_0$, and making the potential equal to the constant $-1/r_1$ for all $r>r_1$. This modified potential has an orbital
which --up to a constant-- is almost identical to the original $n=20$
Rydberg orbital in the region $r<r_0$, but it may now be a {\em
scattering} orbital, behaving very differently in the region $r>r_1$.
We can {\em still} solve Eq.(\ref{ld}) on this scattering orbital,
with $n=20$ on the right-hand side, and find $\mu\simeq-0.441$ at
$r_0$. The altered potential does not have a Rydberg series, yet the
solution of Eq.(\ref{ld}) at $r_0$ can still be interpreted as the
asymptotic quantum defect of the Rydberg series that was lost as a
consequence of the alteration.

The Coulomb tail has nothing to do with the {\em value} of $\mu$. It
only has to do with its definition.

One technical point deserves comment: rather than giving $\mu(r)$
directly, Eq.(\ref{ld}) yields $n^*(r)$, and it would seem that it
needs to be solved with ever increasing accuracy as $n\to\infty$,
since orbitals of different $n$ become essentially identical at large
$n$ in the $[0,r_0]$ region. Only the fractional part of the solution
of Eq.(\ref{ld}) is to be trusted, yielding the fractional
part of $n^*$, which is also equal to the fractional part of the
asymptotic quantum defect, denoted $\{\mu_{\infty}\}$. The integer
part of $\mu_{\infty}$ can be determined by a simple node counting as
$[\mu_{\infty}]=N-N_C$, where $N$ is the number of nodes of the given
orbital in the $[0,r_0]$ interval and $N_C$ that of the correspoding
pure Coulomb orbital. The asymptotic quantum defect $\mu_{\infty}=[\mu_{\infty}]+\{\mu_{\infty}\}$ is thus fully determined this way.

%\begin{figure}
%\begin{center}
%\epsfxsize=80mm \epsfbox{qd_example1_inset.eps}
%\caption{$s$-quantum defect as a function of $n$
%for the potential shown in the inset. The quantum defect is a smooth
%function of $n$ and converges rapidly to its asymptotic value,
%$\mu_{\infty}=-0.441$.} \label{f:qdvsn_example1}
%\end{center}
%\end{figure}

%\begin{figure}
%\epsfxsize=80mm \epsfbox{qdvsn_example1.eps}
%\caption{$s$-quantum defect as a function of $n$ for the potential
%of the first example. The quantum defect is a smooth function of
%$n$ and converges rapidly to its asymptotic value, $\mu=-0.441$.}
%\label{f:qdvsn_example1}
%\end{figure}

%\begin{figure}
%\begin{center}
%\epsfxsize=85mm
%\epsfbox{mu_example1.eps}
%\caption{Solution of Eq.(\ref{ld}) for $\mu$ as a function of $r$, for the potential of Fig.\ref{f:qdvsn_example1}; the logarithmic derivative of the $n=20$ orbital was found numerically as a function of $r$, and at each location $r$, it was inserted into Eq.(\ref{ld}) to get $\mu(r)$.}
%\label{f:qd_example1}
%\end{center}
%\end{figure}

%
%\subsubsection{Potential that yields $\mu=1$}
%As another example, take $\phi_{>r_0}$ to be the $n=7$ Coulomb radial orbital,
%meant to represent (for $r>r_0$) the $n=8$ orbital of some
%potential which differs from $-1/r$ for $r<r_0$, but equals $-1/r$
%for $r\geq r_0$. Applying Eq.(\ref{ld}) for any $r>r_0$ we find
%correctly $\mu=1$, because $n=8$ in the right-hand-side of
%Eq.(\ref{ld}). One of the nodes has been shifted into the region
%$r<r_0$.
%
%\subsection{Results for Helium and Neon}
%\subsubsection*{Helium}

\begin{figure}
\begin{center}
\epsfxsize=80mm
\epsfbox{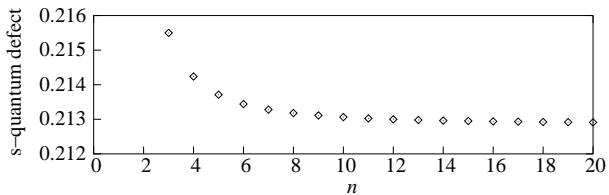}
\caption{$s$-quantum defect as a function of $n$ for the exact KS potential \cite{UG94} of the He atom. The quantum defect converges rapidly to its asymptotic value, $\mu_{\infty}=0.213$.}
\label{f:qdvsn_He}
\end{center}
\end{figure}

\begin{figure}
\begin{center}
\epsfxsize=80mm
\epsfbox{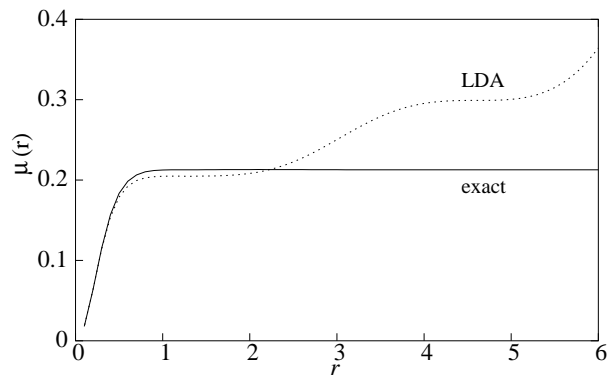}
\caption{He atom: solution of Eq.(\ref{ld}) for $\mu$ as a function of $r$; The $n=20$ orbital was used for the exact case, and the scattering orbital or energy $E=I+\epsilon_{1s}^{LDA}$ was used for the LDA.}
\label{f:He_LDA_and_exact_qd}
\end{center}
\end{figure}

\begin{figure}
\begin{center}
\epsfxsize=80mm
\epsfbox{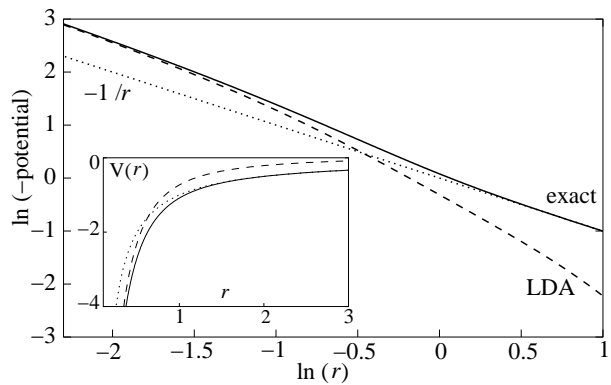}
\caption{Comparison of the exact KS potential of the He atom \cite{UG94} (solid line), and the LDA potential (dashed line). The Coulomb potential is also shown (dotted line). At $r\sim 1$ the exact potential is almost Coulombic. The inset shows the potentials themselves.}
\label{f:He_pot}
\end{center}
\end{figure}

{\underline{\em {Results for Helium and Neon}}}: Consider the first
$s$-Rydberg series of the He atom.  Figure \ref{f:qdvsn_He} shows the
$s$-quantum defect as a function of $n$ from the {\em exact} KS
potential obtained by Umrigar and Gonze \cite{UG94}. The asymptotic
quantum defect $\mu_{\infty}=0.213$ can be extracted through Eq.(\ref{ld}) from
e.g. the $n=20$ orbital, just as it was done in the previous
example. It is clear from the solid line of Fig.\ref{f:He_LDA_and_exact_qd} that at $r_0\sim 1$,
Eq.(\ref{ld}) is already giving an accurate value of $\mu$, which is
quite remarkable considering that the KS potential at $r\sim 1$ is
still not equal to $-1/r$ (see Fig.\ref{f:He_pot}). Most of the
quantum defect is built up close to the nucleus (steep rise for
$0<r<1$ in Fig.\ref{f:He_LDA_and_exact_qd}), and its final value has been reached before the potential becomes purely coulombic.

\begin{figure}
\begin{center}
\epsfxsize=80mm
\epsfbox{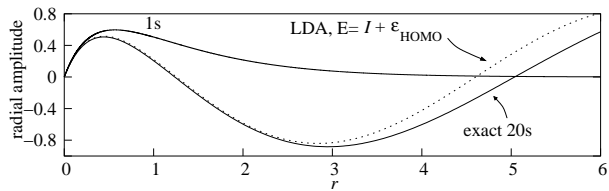}
\caption{Radial orbitals of He: LDA orbital of energy E=0.333 and exact 20s orbital; the HOMO is also shown. The LDA scattering orbital and the exact KS Rydberg orbital are very close in the region $1<r<2$, where the quantum defect can be extracted (see text, and Fig.\ref{f:He_LDA_and_exact_qd}). The fact that the LDA orbital has an incorrect asymptotic behavior at large $r$ is irrelevant for the value of $\mu$, as well as for optical absorption \cite{WMB03}.}
\label{f:orbitals_He}
\end{center}
\end{figure}

The LDA potential runs almost parallel to the exact one in the region
$1<r<2$ (where $\mu_{\infty}$ can already be extracted accurately), and
orbitals corresponding to the same {\em frequency} (exact and LDA) are
therefore very close in that region, see Fig.\ref{f:orbitals_He}. In
the spirit of Ref.\cite{WMB03}, we compare the exact
energy-normalized 20s orbital (which is essentially identical to the zero-energy state
in the region $0<r<6$) and the LDA orbital of energy
$I+\epsilon_{1s}^{LDA}=0.904-0.571=0.333$. Notice how good the LDA
orbital is in the region $1<r<2$. We show in
Fig.\ref{f:He_LDA_and_exact_qd} the solution of Eq.(\ref{ld}) when
this scattering LDA orbital is employed. Clearly, the plateau of the
LDA curve in the $1<r<2$ region is an accurate estimate of the quantum
defect. The value of $\mu$ on this plateau is 0.205, an
underestimation of less than 4\% with respect to the exact value.

Thus, given the ionization potential of the system, LDA gives a very
accurate prediction of the asymptotic quantum defect. The ionization
potential is needed to choose the appropriate LDA scattering orbital,
but the results are not terribly sensitive to it. We repeated the same
procedure with the LDA ionization potential (defined as
$E_{\rm LDA}$(He)$-E_{\rm LDA}$(He$^+$)=0.974) instead of the exact
one, and found $\mu_{\infty}^{\rm LDA}=0.216$,
overestimating the exact $\mu_{\infty}$ by just 1\%.

Our analysis provides a natural way to asymptotically correct the LDA
potential: simply force the LDA plateau of
Fig.\ref{f:He_LDA_and_exact_qd} to stay constant for all $r$. The
resulting function $\mu(r)$ determines a zero-energy orbital which in
turn uniquely determines a potential through inversion of the
Kohn-Sham equations at zero energy. 
%We are currently investigating to
%what extent the resulting potential differs from other
%asymptotically-corrected potentials. 
We emphasize, however, that such
potential is not needed to obtain $\mu_{\infty}$, but the question of
what long-range potential gives rise to the same $\mu_{\infty}$ is
certainly worth addressing. The TH \cite{TH98} or CS \cite{CS00}
methods for asymptotically correcting the LDA potentials require
choosing a radius $r_0$ where the proper tail is to be pasted. Our
analysis also provides a way to rigorously justify such choice, since
$r_0$ should clearly be on the LDA plateau of $\mu(r)$. Minimizing
$\left.d\mu(r)/dr\right|_{r_0}$ determines its precise value of $r_0\sim 1.3$. To test this, we
performed a simple Latter-type asymptotic correction \cite{L55} by
pasting a $-1/r$ tail to the LDA potential shifted downwards by
$v_{\rm LDA}(r_0)-r_0^{-1}$, and scanning through $r_0$. The errors in
both $\mu_{\infty}$ and $\epsilon_{1s}$ were minimized at $r_0\sim
1.5$.

Repeating the same procedure for the Ne atom we found again a
distinctive plateau in the LDA curve of $\mu_{\infty}(r)$ at
$\mu_{\infty}^{\rm LDA}=1.366$, an overestimation of 4\% with respect
to the exact value ($\mu_{\infty}^{\rm exact}=1.313$). 

It has been shown \cite{ARU98} for the case of Ne that 3 numbers are enough to
fit very accurately the entire curve of Kohn-Sham quantum defects of a
given $l$. Although we have only determined one such
number here ($\mu_{\infty}$), we use it nonetheless to approximate all the
$l=0$ KS orbital energies as $\epsilon_{\rm LDA}^{\rm
approx}=-[2(n-\mu_{\infty})]^{-1}$. The results are presented in Table
\ref{t:table2}. The errors are smaller than typical TDDFT errors \cite{RG84}.

%\begin{figure}
%\begin{center}
%\epsfxsize=80mm
%\epsfbox{qd_He.eps}
%\caption{He atom: solution of Eq.(\ref{ld}) for $\mu$ as a function of $r$; The $n=20$ orbital was used}
%\label{f:qd_He}
%\end{center}
%\end{figure}

%\begin{table}
%\caption{\label{t:table2} Transition frequencies and oscillator
%strengths in atomic units for the first six discrete $2p\to ns$
%transitions in Ne, from the exact and LDA KS potentials.}
%\begin{ruledtabular}
%\begin{center}
%\begin{tabular}{c|cc|cc}
%\hline
%trans.&\multicolumn{2}{c}{transition frequency}&
%\multicolumn{2}{c} {oscillator strength}\\
%&LDA&exact
%&LDA&exact\\
%\hline
%2p$\to$3s&0.6052&0.6102&2.38(-2)&2.70(-2)\\
%2p$\to$4s&0.7204&0.7227&3.99(-3)&4.46(-3)\\
%2p$\to$5s&0.7546&0.7556&1.39(-3)&1.57(-3)\\
%2p$\to$6s&0.7692&0.7697&6.49(-4)&7.34(-4)\\
%2p$\to$7s&0.7767&0.7770&3.55(-4)&4.03(-4)\\
%2p$\to$8s&0.7811&0.7813&2.15(-4)&2.45(-4)\\
%\hline
%\end{tabular}
%\end{ruledtabular}
%\end{center}
%\end{table}

\begin{table}
\caption{\label{t:table2} Transition frequencies (in eV) for the first
six discrete $2p\to ns$ transitions in Ne, from the exact and LDA KS
potentials.}
%\begin{ruledtabular}
\begin{center}
\begin{tabular}{c|cc}
\hline
trans.&\multicolumn{2}{c}{transition frequency}\\
&LDA&exact\\
\hline
2p$\to$3s&16.468&16.604\\
2p$\to$4s&19.603&19.666\\
2p$\to$5s&20.534&20.561\\
2p$\to$6s&20.931&20.945\\
2p$\to$7s&21.135&21.143\\
2p$\to$8s&21.255&21.260\\
\hline
\end{tabular}
%\end{ruledtabular}
\end{center}
\end{table}

%\begin{figure}
%\begin{center}
%\epsfxsize=80mm
%\epsfbox{Ne_os.eps}
%\caption{KS optical spectrum of the Ne atom, in atomic units, from the exact and LDA potentials. The discrete part of the spectrum was built with the excitation energies listed in Table \ref{t:table2} and intensities from the photoionization curve of ref.\cite{WMB03}.}
%\label{f:Ne_os}
%\end{center}
%\end{figure}

Finally, our results suggest that the LDA can also be employed to
calculate accurate low-energy electron-ion scattering phase
shifts. In fact, the LDA asymptotic quantum defect found in this work,
$\mu_{\infty}^{\rm LDA}=0.205$, immediately yields, through Seaton's
theorem \cite{S58}, a prediction for the zero-energy s-phase-shift
for electron-He$^+$ scattering: $\delta(E=0)_{\rm
LDA}=\pi\mu_{\infty}^{\rm LDA}=0.644$. The {\em exact} KS phase shift
is $\delta(E=0)_{\rm exact}=0.669$. This value is also remarkably close to
the average of the experimental singlet/triplet zero-energy phase
shifts \cite{WMB05}.

In conclusion, we have shown that rather than modifying the shape of
the LDA potentials one can modify the interpretation of the results of
a {\em pure} DFT-LDA calculation. The results are excellent for the
excitation to high-lying $s$-Rydberg states in He and Ne, and we are working to extend these ideas to Rydberg excitations in molecules.

This work was supported by NSF grant No.CHE-0355405.

\end{document}